\documentclass[prl,aps,showpacs,superscriptaddress,amsmath,amssymb,floatfix,twocolumn]{revtex4}
\usepackage{graphicx}
\usepackage{color}

\begin{document}

\title{Protected quantum computation  with multiple resonators \\ in ultrastrong coupling circuit QED}
\date{\today} 

\author{Pierre Nataf}
\author{Cristiano Ciuti}
\affiliation{Laboratoire Mat\'eriaux et Ph\'enom\`enes Quantiques,
Universit\'e Paris Diderot-Paris 7 et CNRS, \\ B\^atiment Condorcet, 10 rue
Alice Domon et L\'eonie Duquet, 75205 Paris Cedex 13, France}

\begin{abstract}
We investigate theoretically the dynamical behavior of a qubit obtained with the two ground eigenstates of an ultrastrong coupling circuit-QED system consisting of 
a finite number of Josephson fluxonium atoms inductively coupled to a transmission line resonator.  We show an universal set of quantum gates by using multiple transmission line resonators (each resonator represents a single qubit). We discuss the intrinsic 'anisotropic' nature of noise sources for fluxonium artificial atoms. Through a master equation treatment  with  colored noise and manylevel dynamics, we prove that, for a general class of anisotropic noise sources, the  coherence time of the qubit and the fidelity of the quantum operations can be dramatically improved in an optimal regime of ultrastrong coupling, where the ground state is an entangled photonic 'cat' state.
     \end{abstract}
     
     \pacs{03.65.Yz; 85.25.Hv; 42.50.Pq; 03.67.Pp}
\maketitle

The study of quantum decoherence is believed to be crucial in order to understand the transition from the microscopic quantum world to the macroscopic classical one. Moreover, a control and limitation of decoherence is essential towards the realization of a robust, scalable quantum computer. The study of cavity QED systems in atomic physics \cite{brune} has led to spectacular fundamental investigations of non-unitary evolution due to decoherence mechanisms. In particular, it has been possible to observe the fragility of  states of the form $\vert\Psi_{cat}\rangle = \frac{1}{\sqrt{2}} \{ \vert \alpha  \rangle_{phot} \vert g \rangle_{at} +  \vert\alpha e^{i\eta} \rangle_{phot} \vert e \rangle_{at}\}$ (usually dubbed 'cat states'\cite{brune}), where $\vert \alpha  \rangle_{phot}$ is a coherent photon state with a large mean photon number $\vert \alpha \vert^2 \gg 1$,   $\vert \alpha e^{i\eta} \rangle_{phot}$ is another coherent state with a phase difference $\eta$, while $\vert g \rangle_{at}$ ($\vert e \rangle_{at}$) is the ground (excited) state of a two-level atom \cite{brunesch,martinis1}. These states have been prepared in a cavity QED system well described by the Jaynes-Cummings model, where the ground state is  $\vert0  \rangle_{phot} \vert g \rangle_{at}$, i.e., the vacuum of photons times the atomic ground state.  
Recently, a growing interest has been generated by the so-called ultrastrong coupling regime of cavity (circuit) QED , both theoretically \cite{ciuti,ciutiPRA2006,devoretstrong, Bourassa, degvacua,pedro,nori,milburn,theo} and experimentally \cite{anappara,gunter,niemczyk,yanko,fedorov}. Such a regime is achieved when the vacuum Rabi frequency $\Omega_0$, which quantifies the coupling between one photon and one elementary matter excitation, is comparable or larger than the cavity (resonator) photon frequency $\mathbf{\omega}_{cav}$ . 
In such a regime, the Jaynes-Cummings model based on the rotating wave-approximation (valid for small ratio  $\Omega_0/\omega_{cav}$) breaks down. In particular,  the ground state of the system is no longer the standard vacuum: recently, it was shown \cite{degvacua,nori} that  in the limit of very large coupling the ground state can become quasi-degenerate with the entangled structure:\begin{equation}
\label{ground}
\vert \Psi_G  \rangle \simeq \frac{1}{\sqrt{2}} \left (\vert \alpha \rangle_{ph}\, \Pi_{j=1}^{N} \vert + \rangle_j + 
(-1)^N\vert -\alpha \rangle_{ph}\, \Pi_{j=1}^{N} \vert - \rangle_j 
\right )
\end{equation}
where $N$ is the number of atoms embedded in the cavity resonator, $|\alpha \rangle_{ph}$ is a coherent state for the photonic field which satisfies $|\alpha|\sim\sqrt{N} \Omega_0/\omega_{cav}$,  and $\vert \pm \rangle_j$ are  pseudo-spin polarized states  for the $j$-th artificial  atom, which are defined in the following. For each two-level system $\{ \vert e\rangle_j, \vert g \rangle_j\}$ one can introduce the Pauli operators $\hat{\sigma}_{x}^j = \vert e\rangle_j \langle g \vert_j +   \vert g\rangle_j \langle e \vert_j$,  $\hat{\sigma}_{y}^j = i(\vert g\rangle_j \langle e \vert_j -   \vert e\rangle_j \langle g \vert_j)$ and $\hat{\sigma}_{z}^j = (2 \vert e\rangle_j \langle e \vert_j -1)$.
With a  light-matter coupling Hamiltonian of the form $H_{coupling}=\sum_{j=1}^N \lambda_j \hat{a} \, \hat{\sigma}_x^j+\rm{h.c}$, ($\lambda_j$ being the local  coupling strength and $\hat{a}$ the photonic bosonic annihilation operator),   $\vert \pm \rangle_j = \frac{1}{\sqrt{2}} (\vert e \rangle_j \pm \vert g \rangle_j)$ are the eigenstates of $\hat{\sigma}_x^j$. 
Interestingly, the (orthogonal) first excited state has  the similar form:
\begin{equation}
\label{excited}
\vert \Psi_E \rangle \simeq \frac{1}{\sqrt{2}} \left (\vert \alpha \rangle_{ph}\, \Pi_{j=1}^{N} \vert + \rangle_j  -(-1)^N
\vert-\alpha \rangle_{ph}\, \Pi_{j=1}^{N} \vert - \rangle_j 
\right )
\end{equation}
In this letter, we show how $\vert \Psi_G\rangle$ and $\vert \Psi_E \rangle$ surprisingly can form a robust qubit, whose decoherence can diminish while increasing the `size' of the corresponding photonic `cat' states (see Fig. 1). Moreover, we also provide a universal set of quantum computation gates and demonstrate via a thorough master equation treatment the fidelity enhancement in a regime of ultrastrong coupling. 
  \begin{center}
  \begin{figure}[h!]
\includegraphics[width=210pt]{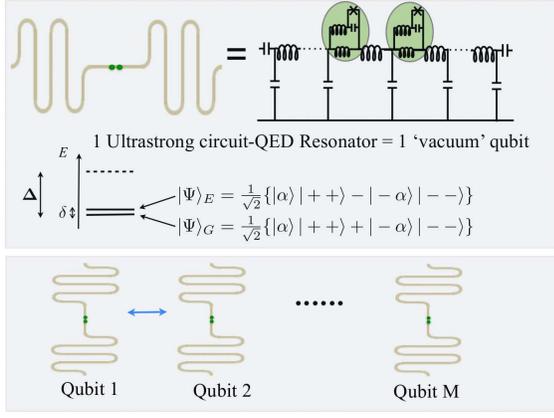}
\caption{\label{sketch} Description of the considered system. The building block is a superconducting transmission line resonator embedding $N$ Josephson atoms ($N=2$
in the sketch here).
 By choosing judiciously the type of artificial  atom (the depicted circuit represents fluxonium atoms  inductively coupled to the resonator) , the first two ground levels of the resonator are  entangled states ($|\alpha\rangle$ is a photon coherent state, $\vert \pm\rangle$ is a Josephson junction state 'polarized' along the pseudospin $x$-direction). One resonator represents a single qubit:  a register of $M$ qubits is given by $M$ resonators.}
\end{figure}
 \end{center}
The energy difference $\delta$ between the two considered states diminishes exponentially\cite{degvacua,nori} with the vacuum Rabi coupling, namely
$\delta \sim \omega_{eg} \exp(-2 \frac{\Omega_0^2}{\omega_{cav}^2} N)$, where $\omega_{eg}$ is the frequency of the single atom two-level transition (set to be equal
to the cavity mode frequency). Either in the ultrastrong ($\Omega_0/\omega_{eg} \to +  \infty$) or `thermodynamic' ($N \to + \infty$) limit, the two states become degenerate. 
In the ultrastrong coupling limit, the other excited states are much higher in energy, separated by a frequency gap $\Delta \sim \omega_{eg} >> \delta$. 
Importantly, these interesting features can not be obtained in every ultrastrong light-matter coupled system. In particular, using the Pauli matrix language, the `direction' of the bare atomic Hamiltonian must be orthogonal to the one of the light-matter interaction Hamiltonian\cite{nori}. This is the case in the following spin-boson Hamiltonian:
\begin{eqnarray}
\label{ham}
\hat{H}/\hbar  =\omega_{cav} \hat{a}^{\dag}\hat{a} \,+\, \frac{\omega_{eg}}{2}\sum_{j=1}^{N} \hat{\sigma}_{z}^j + \sum_{j=1}^N i\frac{\Omega_0}{\sqrt{N}}(\hat{a}-\hat{a}^{\dag}) \hat{\sigma}_x^j.
\end{eqnarray}
In the present letter, we limit our description to a single bososic mode  and an uniform light-matter coupling , but all the following results may be generalized to several and spatially non uniform modes{\cite{degvacua}}.
 
 It has also been shown recently that in the ultrastrong coupling regime, the quasi-degeneracy of the states $\vert \Psi_G  \rangle $ and  $\vert \Psi_E  \rangle$ is robust with respect to a local and static perturbation of the type $
H^{pert}_{y,z} = \sum_{j=1}^{N} h_{y,j} \hat{\sigma}_{y,j} +  h_{z,j} \hat{\sigma}_{z,j} $
where $h_{y,j}$ and $h_{z,j}$ are random perturbation amplitudes \cite{degvacua}.
The reason is that in the subspace $\{ \vert \Psi_G \rangle , \vert \Psi_E\rangle \}$ such perturbation couples (at the $N^{th}$order) coherent states of opposite phase $|-\alpha\rangle$ and $|\alpha\rangle$.
The effect of the perturbation is proportional to the overlap $\langle -\alpha | \alpha \rangle = \exp{(-2 |\alpha|^2)} \sim \exp(-2 \frac{\Omega_0^2}{\omega_{cav}^2} N)$.
  Indeed, the stronger is  the coupling $\Omega_0$ or the larger is the number of artificial atoms $N$, the larger
is $\vert \alpha \vert^2$, the `size'  of the photonic 'cat' states  $\vert \Psi_G \rangle$ and $\vert \Psi_E \rangle$.
Importantly, the protection is not complete\cite{Doucot}, because these states are not robust with respect to noise terms like $H^{pert}_{x} =  \sum_{j=1}^{N} h_{x,j} \hat{\sigma}_{x,j}$ and  $H^{pert}_{\hat{a}} = h_{a} \hat{a} + h_{a}^*\hat{a}^{\dagger}$, namely the noise in the direction of the light-matter coupling and the noise associated to the resonator field.
\,However, if in a superconducting system, perturbations like $H^{pert}_{y,z}$ happen to be the dominant ones, the lifetime and the fidelity of the quantum operation involving the states $\vert \Psi_G \rangle$ and $\vert \Psi_E \rangle$ can  be dramatically improved by increasing $\Omega_0/\omega_{eg}$ and/or $N$.

\begin{center}
  \begin{figure}[h!]
\includegraphics[width=250pt]{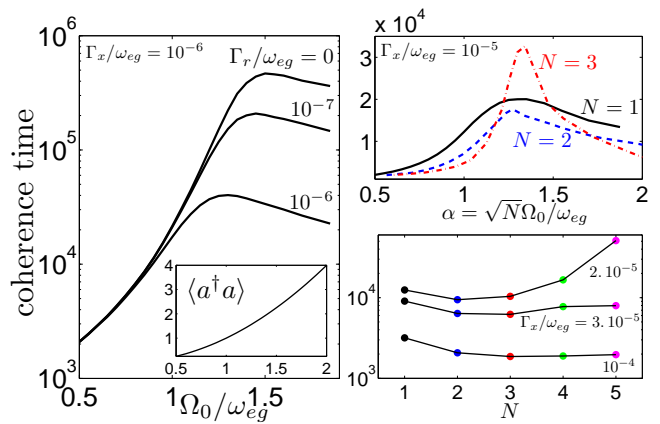}
\caption{ \label{coherence} Coherence time in units of $1/\omega_{eg}$  calculated via the master equation (\ref{mast})  for $\omega_{eg}=\omega_{cav}$  and with the initial state $|\Psi_0\rangle=\cos(\theta) |\Psi_E\rangle\,+\,\sin(\theta) e^{i\phi}|\Psi_G\rangle$. Results are averaged over the possible initial values for $\theta$ and  $\phi$. 
Left panel: coherence time versus the normalized vacuum Rabi frequency $\Omega_0/\omega_{eg}$ for one atom ($N=1$) with Josephson loss rates  $\{\Gamma_x,\Gamma_y,\Gamma_z\}= \omega_{eg} \{10^{-6},10^{-3},10^{-3}\}$. The  different cavity loss rates: $\Gamma_{r}/\omega_{eg}=10^{-6},10^{-7},0$ correspond \cite{brune} to different quality factors $Q=\omega_{eg}/(4\pi\Gamma_{r})\simeq 10^5,10^6,\infty$.
 Inset : the number of photons $\langle n\rangle=|\alpha|^2=\langle a^{\dag}a\rangle$ is plotted versus  $\Omega_0/\omega_{eg}$ for $N=1$.
Top right panel: coherence time for $N=1$, $2$ and $3$ atoms for $\Gamma_{r}/\omega_{eg}=10^{-6}$ and with a lower anisotropy in the atomic loss rates: $\{\Gamma_x,\Gamma_y,\Gamma_z\}= \omega_{eg} \{10^{-5},10^{-3},10^{-3}\}$ as a function of the photonic amplitude $\alpha=\sqrt{N}\Omega_0/\omega_{eg}$.  Bottom right panel: maximum coherence time as a function of the number of atoms $N$ for different values of $\Gamma_x$, hence for different noise anisotropy. }
\end{figure}
 \end{center}

In fact, among the different flux Josephson atoms \cite{mooij,niemczyk,fedorov,Devoret}, this noise anisotropy appears to be realistic at least for a fluxonium\cite{Devoret} . Under conditions detailled in \cite{Devoret}, its Hamiltonian can be written as:
\begin{eqnarray}
\label{Hflux}
H_{F}= 4E_{C_J} \hat{N}_J^2+E_{L_J}\frac{(\hat{\varphi}_{J})^2}{2} - E_J\cos(\hat{\varphi}_{J}+\Phi_{ext})
\end{eqnarray}
The Hamiltonian parameters are subject to noise fluctuations :  $\Phi_{ext} = \pi +\Delta \Phi_{ext}$ with $\Delta \Phi_{ext}$ some flux noise (in units of $\Phi_0=\hbar/2e$), 
$ E_J= E_J+\Delta E_J $ with $\Delta E_J = \Delta I_0/ \Phi_0$ proportional to the critical current fluctuation,   $\hat{N}_J=\hat{N}_J+\Delta N_0$, $\Delta N_0$ being the charge offset fluctuation. One can also introduce
some capacitive and inductive noise $E_{C_J}=E_{C_J}+\Delta E_{C_J}$ and $E_{L_J}=E_{L_J}+\Delta E_{L_J}$.
When the fluctuation sources are off, the first two eigenstates of the fluxonium are very well isolated from the higher states provided that $E_J \gg E_{L_J}$ and $E_J\gg E_{C_J}$. Then, the Hamiltonian (\ref{Hflux}) reads $\hat{H}_{F} \simeq \hbar (\omega_{eg}/2) \hat{\sigma}_z$  in the basis of the two first eigenstates which are symmetric and antisymmetric superpositions of  clockwise and anticlockwise persistent current states. On the same basis $\hat{\varphi}_{J} \simeq -\varphi_{01} \hat{\sigma}_x$ and $\hat{N}_J \simeq \frac {\omega_{eg}}{8E_C}\varphi_{01}\hat{\sigma}_y$ (where $\varphi_{01}\simeq$ 3).
The fluctuations produce (at the first order) the perturbation:
\begin{eqnarray}
\label{Hfluxpert}
\hat{H}_{F, pert}/\hbar \simeq \ \Delta \Phi_{ext} sin(\varphi_{01}) (E_J/\hbar) \hat{\sigma}_x+\Delta N_0\varphi_{01} \omega_{eg} \hat{\sigma}_y \,\,\,\,\, \\\nonumber 
+\,(\frac{\partial\omega_{eg}}{\partial E_J} E_J\frac{\Delta I_0}{I_0} +\frac{\partial\omega_{eg}}{\partial E_{C_J}} E_{C_J} \frac{\Delta E_{C_J}}{E_{C_J}} + \frac{\partial\omega_{eg}}{\partial E_{L}} E_L \frac{\Delta E_L}{E_L})\hat{\sigma}_z
\end{eqnarray}
The spectral density of the flux noise is typically  $S_{\Delta \Phi_{ext}}^{1/2} \approx 10^{-6}/\sqrt{Hz}$ \cite{fluxnoiseth,fluxnoiseyo}. The critical current noise $\Delta I_0/I_0=\Delta E_J/E_J$, which is also believed to follow a $1/f$ law \cite{harlingen,ithier}, has been recently measured\cite{arxivdevo} in a fluxonium : $S_{\Delta E_J/E_J}^{1/2}\approx 3.10^{-5}/\sqrt{Hz}$. It proves that the dissipation due to the $\hat{\sigma}_z$ channel is much larger than the $\hat{\sigma}_x$ channel contribution.
  To study the behavior of the qubit $\{\vert \Psi_G \rangle,\vert \Psi_E \rangle\}$ in the presence of dissipation, we used the master equation \cite{breuer}:
    \begin{eqnarray}
\label{mast} 
\frac{d\hat{\rho}}{dt}=\frac{1}{i\hbar}[\hat{H},\hat{\rho}]\,+\sum_{r=r_v,r_f}\hat{U}_{r} \hat{\rho} \hat{S}_{r}  + \hat{S}_{r} \hat{\rho} \hat{U}_{r}^{\dag} - \hat{S}_{r}\hat{U}_{r}\hat{\rho} - \hat{\rho} \hat{U}_{r}^{\dag}  \hat{S}_{r}
\nonumber\\
    +\,\sum_{j=1}^N\sum_{m=x_j,y_j,z_j}  \hat{U}_m\hat{\rho} \hat{S}_m  + \hat{S}_m \hat{\rho} \hat{U}_m^{\dag} - \hat{S}_m \hat{U}_m \hat{\rho} - \hat{\rho} \hat{U}_m^{\dag}  \hat{S}_m\,\,\,\,\,\,\,\,\, 
    \end{eqnarray}
         where $\hat{\rho}$ is the density matrix,   $\hat{H}$ refers to Hamiltonian (\ref{ham}) and where the `jump' operators are
  $\hat{S}_{r_v} =\hat{a}+\hat{a}^{\dag}$, $\hat{S}_{r_f} =i(\hat{a}-\hat{a}^{\dag})$, $\hat{S}_{x_j}=\hat{\sigma}_x^j$, $\hat{S}_{y_j}=\hat{\sigma}_y^j$, $\hat{S}_{z_j}=\hat{\sigma}_z^j$. 
  Moreover\cite{timenote},
 \begin{eqnarray}
 \label{nuk}
  \hat{U}_k =\int_0^{\infty} \nu_k(\tau) e^{-\frac{i}{\hbar}\hat{H} \tau} \hat{S}_{k} e^{\frac{i}{\hbar}\hat{H} \tau} d\tau ,\\
  \nu_k(\tau) =\int _{-\infty}^{\infty} \Gamma_k (\omega) \{n_k(\omega) e^{i \omega \tau}\,+\,[n_k(\omega)+1]  e^{-i \omega \tau}  \} d\omega \nonumber,
      \end{eqnarray}
for $k=r_v, r_f$ or $k=x_j,y_j,z_j$ $\forall j=1..N$.\\
Here we consider the zero temperature limit\cite{temperature}, where
 the spectral functions $\Gamma_k(\omega)$ must vanish for $\omega<0$  because they are proportional to the density of states (of the baths) at energy $\hbar \omega$.
For sake of simplicity, we have set $\Gamma_k(\omega)=\Gamma_k$ for $ \omega \in [0; \omega_c ]$ and  $\Gamma_k(\omega)=0$ elsewhere $\forall  k$,  with $\omega_c$  an upper cut-off which is consistent with decreasing spectral noise. Finally, one must include many excited states in the master equation treatment.
To investigate the robustness of the coherence between the  2 quasi-degenerate vacua $|\Psi_G \rangle$ and $| \Psi_E \rangle$ 
, we have studied the non-unitary dynamics of the initially prepared pure state $|\Psi_0\rangle=\cos(\theta) |\Psi_E\rangle\,+\,\sin(\theta) e^{i\phi}|\Psi_G\rangle$ in presence of anisotropic Josephson dissipation rates $\Gamma_y,\Gamma_z \gg \Gamma_x$  and for several cavity loss rates $\Gamma_{r}/\omega_{eg} = \Gamma_{r_v}/\omega_{eg}=\Gamma_{r_f}/\omega_{eg}$ (see caption of Fig. 2).  
Our simulations plotted in Fig. \ref{coherence} prove that the coherence time increases while  increasing the normalized vacuum Rabi
frequency $\Omega_0/\omega_{eg}$. Indeed, if the dominant dissipation channels are along the $y$ and $z$ directions, their effect decreases as $\exp{(-2 |\alpha|^2)}$ where $\alpha=\sqrt{N}\Omega_0/\omega_{cav}$. Hence,  the coherence time is enhanced exponentially  before reaching a saturation value given by $\Gamma_r$, $\Gamma_x$ and eventually  decreasing with the usual power law of  cat states. The location of the coherence time peaks with respect to the photonic amplitude $\alpha=\sqrt{N}\Omega_0/\omega_{eg}$ is almost independent of the number of atoms $N$ (see top right panel of Fig. 2), indicating that $\alpha$ is the relevant dimensionless parameter for the protection. Depending on $\Gamma_r$ and $\Gamma_x$ (see bottom right panel of Fig. 2), the maximum coherent times have a different behavior versus $N$. For smaller values of $\Gamma_x$, the protection increases monotonically with $N\geq2$  (we have been able to calculate up to $N=5$). For larger values of $\Gamma_x$ instead the maximum of the coherence time is achieved for $N=1$. 
Finally, since the number of photons $\langle n\rangle$ of $|\Psi_G \rangle$ and $| \Psi_E \rangle$ increases like  $\frac{\Omega_0^2}{\omega_{cav}^2} N$ (see Fig. \ref{coherence}), we conclude that there is a regime where the larger is the number of photons in  $|\Psi_G \rangle$ and $| \Psi_E \rangle$, the stronger is their robustness against decoherence contrary to the usual cavity QED `cat' states \cite{brune}, obtained when $\Omega_0/\omega_{eg}<<1$. Indeed, it is well known\cite{brune,brunesch,martinis1,martinis2} that the coherence time of those standard 'cat' states decreases monotonically while increasing their size. 

Now, we show how to obtain an universal set of gates for quantum computation \cite{univ} using the two states $\vert \Psi_G\rangle$ and  $\vert \Psi_E\rangle$ as computational basis for the qubit and we will study the fidelity of such quantum operations. 
    One begins by showing how to  get the dynamical gate   $e^{-i \theta_x \hat{\Sigma}_x}$ in the basis $\vert \Psi_G\rangle$ and  $\vert \Psi_E\rangle$,
   where $\hat{\Sigma}_x = \vert \Psi_G\rangle \langle \Psi_E\vert +  \vert \Psi_E\rangle \langle \Psi_G\vert$ is the $x$-direction Pauli matrix associated to this (collective) vacuum qubit. To do so, one  can add a coupling between the flux of one Josephson atom embedded into the resonator (for instance the first atom)  and  an external, classical and tunable magnetic field $\Phi_s(t)$. This leads to an additional Hamiltonian term of the type $M\Phi_s(t) \hat{\varphi}_j^1= C(t) \hat{\sigma}_x^1$ where $\hat{\varphi}_j^1$ is the flux across the Josephson junction of the first artificial atom. 
Such perturbation lifts the degeneracy  of the fundamental subspace so that the new two first eigenstates are $|+\rangle|+\alpha\rangle$ and  $|-\rangle|-\alpha\rangle $ with a splitting $\delta(t)=2C(t)$ and where we have replaced $\Pi_{j=1}^{N} \vert \pm \rangle_j$ by
$|\pm\rangle$ to simplify the notation.
By adiabatically shaping the time-dependence of $C(t)$ it is possible to create a dynamical gate $e^{-i \theta_x \hat{\Sigma}_x}$ with
$\theta_x= \int_0^T C(t) dt$ with [0;T] the gate time interval.
  
Now, we show how to get a second single-qubit gate, namely $e^{-i \theta_z \hat{\Sigma}_z}$ where  $\hat{\Sigma}_z = 2 \vert \Psi_E\rangle \langle \Psi_E\vert -1$.   $\vert \Psi_G\rangle$ and
  $\vert \Psi_E\rangle$ have an energy splitting $\delta$ exponentially decreasing as a function of $\Omega_0$. By modulating in time $\Omega_{0}$, one gets the desired quantum gate. Acting adiabatically,  the rotation angle will be $\theta_z= \int_0^T  \delta(t) dt = \int_0^T  \delta(\Omega_0(t)) dt$.  
Even without  optimizing the temporal shape of $t \rightarrow \Omega_0(t)$ , excellent fidelities can be reached. For instance, for the Z-Pauli gate (corresponding to $\theta_z=\pi/2$), with one atom and for a linear back and forth  between $\Omega_0/\omega_{eg}=2$ and $\Omega_0/\omega_{eg}=1.3$, fidelity  $\geq 99.9\%$  is obtained  for a typical time $T \sim 300/\omega_{eg}$ in presence of realistic dissipation. In practice, to modulate in situ $\Omega_0(t)$, one can use an intermediate loop between the resonator and the artificial atom  with a tunable magnetic flux through it \cite{degvacua,pedro}.
  
  \begin{center}
  \begin{figure}[h]
\includegraphics[width=230pt]{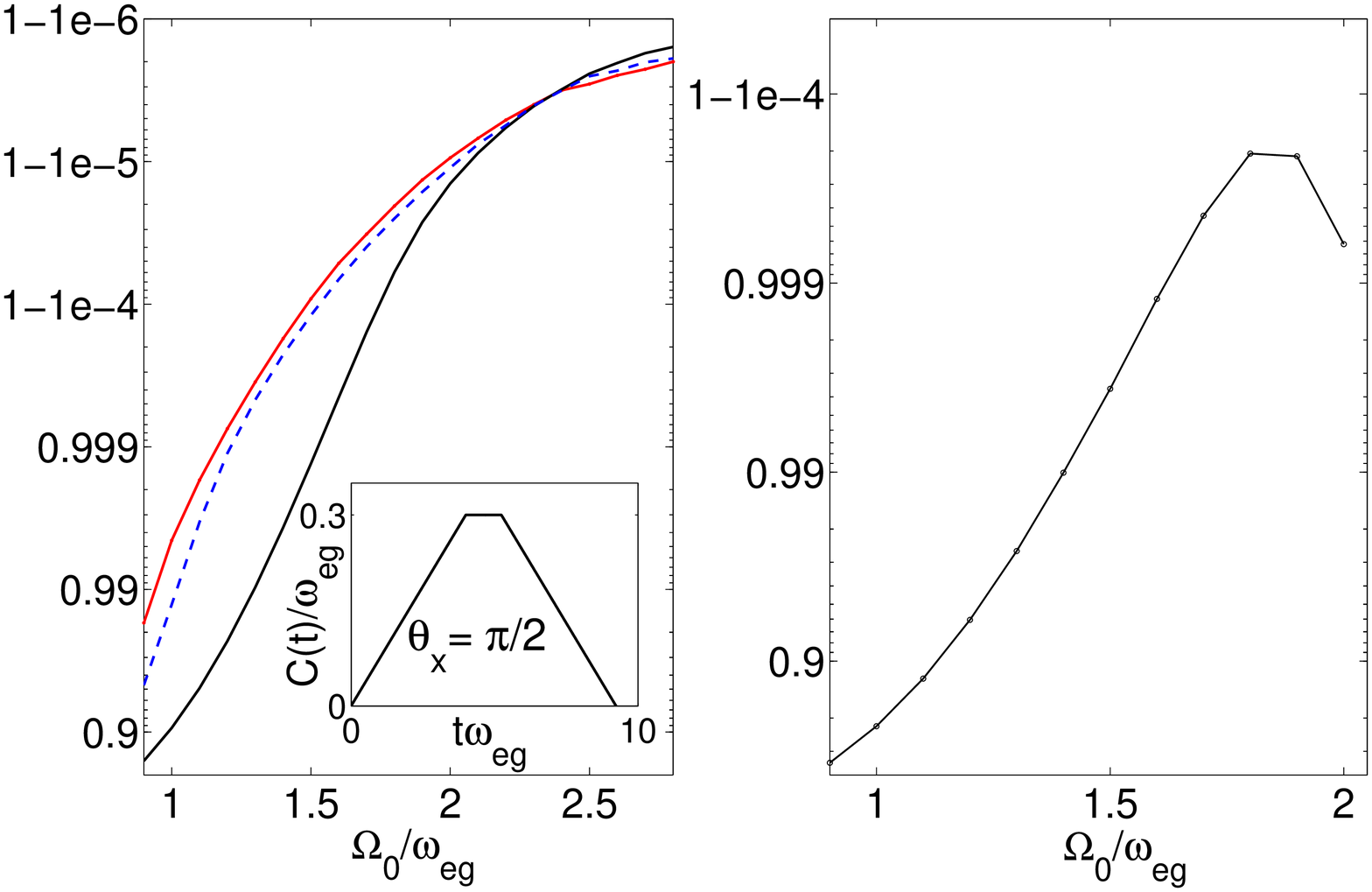}
\caption{\label{rabiN0N1N2N3} Left panel: fidelity of the single qubit X rotation gate (for $\theta_x=\pi/2 $) versus $\Omega_0/\omega_{eg}$. Master equation (\ref{mast}) was used with a time-dependent Hamiltonian, $\hat{H}(t)=\hat{H}(t=0) +C(t)\hat{\sigma}_x^1$ with $\hat{H}(t=0)$ the initial spin-boson Hamiltonian (\ref{ham}) with $N=1$ (black solid), $N=2$ (blue dashed) and $N=3$ (red solid) Josephson atoms in the resonator.  Inset: time evolution of $C(t)$. Right panel: fidelities of the 2-qubit gate  $e^{-i\theta_{x_{12}} \hat{\Sigma}_{x_1} \otimes  \hat{\Sigma}_{x_2}}$  for $\theta_{x_{12}}=\pi/2$ with respect to  $\frac{\Omega_0}{\omega_{eg}}$ the vacuum Rabi Frequency in each resonator in which there is  1 atom embedded.  
The 2-qubits coupling constant $C^{12}(t)$ follows the same time evolution as $C(t)$ in the inset. Note that we have included noise in the mutual coupling, via the jump operator  $\hat{S}_{x12}= \hat{\sigma}^{1}_{x,1}\hat{\sigma}^{1}_{x,2}$ and the loss rate  $\Gamma_{x12}= \omega_{eg} 10^{-6}$.}
\end{figure}
 \end{center}
 
 In order to get a complete set of quantum operations, one needs to perform  a 2-qubit control gate.  Here, we will describe how to obtain the conditional quantum gate
 $e^{-i\theta_{x_{12}} \hat{\Sigma}_{x_1} \otimes  \hat{\Sigma}_{x_2}} $  in the 4-dimensional basis $\{\{|\Psi_G\rangle_1 ,|\Psi_E\rangle_1\}\otimes\{|\Psi_G\rangle_2 ,|\Psi_E\rangle_2 \}\} = \{\frac{1}{\sqrt{2}}(|+\rangle|+\alpha\rangle_1 \pm |-\rangle|-\alpha\rangle_1 )\otimes \frac{1}{\sqrt{2}} (|+\rangle|+\alpha\rangle_2 \pm |-\rangle|-\alpha\rangle_2 )\} $ where 1  (2) stands for the resonator number. 
For our goal, one way 
 is provided by  a direct magnetic mutual coupling\cite{mutual} $M^{12}(t)\hat{\varphi}_j^1 \hat{\varphi}_j^2 $, between 2 fluxonium atoms (one in each resonator), giving the Hamiltonian $
 \hat{H}_{12}=\hat{H}_{1}+\hat{H}_{2}+C^{12}(t) \hat{\sigma}^{1}_{x,1}\hat{\sigma}^{1}_{x,2}
$, where $\hat{H}_{1}$ ($\hat{H}_{2}$) stands for the spin-boson Hamiltonian (\ref{ham}) for the resonator 1 (2), while  $\hat{\sigma}^{1}_{x,1}$ (resp. $\hat{\sigma}^{1}_{x,2}$) stands for the x-Pauli matrix acting on the first two levels system of the resonator 1 (2).
 Applying such a perturbation will partially lift the 4 times degeneracy of the fundamental subspace so that the two states  ($|+\rangle|+\alpha\rangle_1 \otimes |+\rangle|+\alpha\rangle_2$ and  $|-\rangle|-\alpha\rangle_1 \otimes |-\rangle|-\alpha\rangle_2$) will have a different energy than the states ($|+\rangle|+\alpha\rangle_1 \otimes |-\rangle|-\alpha\rangle_2$ and  $|-\rangle|-\alpha\rangle_1 \otimes |+\rangle|+\alpha\rangle_2$). 
 Fidelity of that operation for $\theta_{x_{12}}=\pi/2$ is given in the right panel of  Fig. \ref{rabiN0N1N2N3}  in presence of dissipation, showing again the enhancement  for increasing
 values of the normalized vacuum Rabi frequency. 
Other proposals for the practical coupling between the 2 resonators could be envisaged\cite{2fluxqubits,doublesol}. Concerning the read-out of our qubit, this can be done  by a projective measurement on the states $|+\rangle|+\alpha\rangle$ and  $|-\rangle|-\alpha\rangle $:  the flux across the Josephson junctions is polarized and can be in principle measured via the surrounding quasi-static magnetic field.  
 
In conclusion, we have shown that it possible to considerably enhance the coherence times of a  qubit given by the first two eigenstates of a  circuit QED system in the ultrastrong coupling regime:  such states are entangled states of photons and polarized Josephson atomic states, which are robust with respect to a  general class of 'anisotropic' environment.  In our proposal, the resonator is used to protect quantum information\cite{protect,ibm} , contrary to the approach \cite{blais,maier,dicarlo1} where it acts as a bus  joining several embedded Josephson qubits. The present work shows that the qualitative modification of the quantum ground state in ultrastrong coupling circuit QED can have a significant impact on the decoherence and manipulation of quantum states in multiple resonators.      
We would like to thank M.H.Devoret for a critical reading of the manuscript and useful discussions.\\


\begin{thebibliography}{}
\bibitem{timenote}
       The Hamiltonians $\hat{H}$ considered in the numerical simulations are either time-independent (for the calculation of the coherence times), either slowly-varying (for the calculation of the quantum gate dynamics) within a typical time scale much longer than the reservoir correlation time ($\sim 1/\omega_c$). In the latter case, $\hat{U}(t)$ is time-dependent, its expression being approximated by adiabatically inserting $\hat{H}(t)$ in Eq. (7).
\bibitem{temperature}To have a splitting $\delta \gg k_B T$ (where $T$ is the bath temperature, around $20$ mK in dilution fridges) one has to increase  $\omega_{eg}$  by scaling $E_J$, $E_C$, and $E_L$. In the interesting range of parameters of  the numerical solutions, $\delta$ is between 1 or 2 orders of magnitude smaller than $\omega_{eg}$.
\bibitem{protect} In Ref. \cite{ibm}, the oscillator is not used as a quantum bus, but as a stabilizer for the weakly coupled Josephson atom.
\end{thebibliography}
  \end{document}